\documentstyle[12pt]{article}
\begin{document}

\begin{flushright}
INFNNA-IV-2000/5
\end{flushright}
\vspace*{1.cm}

%%%%%%%%%%%%%%%%%%%%%%%%%%%%%%%%%%%%%%%%%%%%%%%%%%%%%%%%%%%%%%%%%%%%%%%%%%%%%%%%%%%%%%%%%%%%%%%%%%%%%%%%%%%%%%%%%%%%%%%%%%%%%%

\begin{center}{Theory of rare kaon decays\footnote{\small Invited 
plenary talk at the 3rd International Conference on B Physics and CP Violation (BCONF99), Taipei,
Taiwan, 3-7 Dec 1999.}}\\
\vspace*{1.6cm}
{Giancarlo D'Ambrosio \\
\vspace*{0.4cm}
{\em INFN-Sezione di Napoli, 80126 Napoli Italy}}
\vspace*{1.6cm}
\begin{abstract}
We review some recent theoretical results on rare kaon decays. Particular
attention is devoted to find Standard Model tests. This is theoretically
easy in $K\rightarrow \pi \nu \overline{\nu }$, while a careful study of the
long distance contributions is needed for $K_{L}\rightarrow \pi ^{0}e%
\overline{e}$, $K_{L}\rightarrow \mu \overline{\mu }$ and $K\rightarrow \pi
\pi \gamma$ .
\end{abstract}
\end{center}

\section{Introduction}

Historically, rare kaon decays, like $K_{L,S}\rightarrow \mu \overline{\mu }%
, $ $K\rightarrow \pi \nu \overline{\nu },$ and $K^{0}-\overline{K^{0}}$
mixing have been fundamental to understand weak interactions and particle
physics: i) the solution to the large Flavour Changing Neutral Currents
(FCNC) was GIM\ mechanism and ii) CP violation in the Standard Model (SM)
led to three families \cite{reviews,DI98}.\ Originally QCD$\;$corrections to 
$W-$ box, $Z^{0}-$ and $\gamma -$penguin contributions were not evaluated
and phenomenological models were used to estimate long distance
contributions \cite{GL}. For instance for $K_{S}\rightarrow \gamma \gamma ,$
since the short distance amplitude was vanishing, the long distance one was
computed by a $\pi ^{+}-\pi ^{-}\ $ loop with a $K_{S}\pi \pi $ constant
weak vertex from $A(K_{S}\rightarrow \pi \pi $)$_{{\rm {exp}}}$ and $\gamma
^{\prime }$s coupled electromagnetically; $A(K_{S}\rightarrow \gamma \gamma $%
) then is finite \cite{GL} but with at least two legittimate questions: i)
what is the correct off-shell extrapolation for $K_{S}\rightarrow \pi \pi $
and ii) why $K_{S}F^{\mu \nu }F_{\mu \nu }$ local operators (and higher
dimensional operators like $\partial ^{2}K_{S}F^{\mu \nu }F_{\mu \nu }$ )
are suppressed?

There have been two major theoretical steps in this field since the seminal
paper of Gaillard-Lee: i) the OPE (Operator Product Expansion) and ii)
Chiral pertubation theory ($\chi PT).$ In OPE the physical processes are
determined by an effective hamiltonian, written as a product of local
operators $O_{i}$ and (Wilson) coefficients $c_{i}:$ ${\cal H}
_{eff}=\sum_{i} $ $c_{i}(\mu )$ $O_{i}(\mu );$ indeed the scale dependence
must cancel in the product since physical processes are $\mu -$independent.
Thus OPE factorizes short ($c_{i}(\mu )$ ) and long distance contributions.
Long distance matrix elements are evaluated by symmetry arguments, i.e. $%
\chi PT$ \cite{Weinberg1,DI98}. Then, for instance, it is easy to understand
the problems posed above for $K_{S}\rightarrow \gamma \gamma $: chiral
symmetry constrains the momentum dependence of the vertex $K_{S}\pi \pi $
and imposes the ${\cal O}(p^{4})$ $K_{S}F^{\mu \nu }F_{\mu \nu }$ local
operator to be zero, while chiral power counting suppresses $\partial
^{2}K_{S}F^{\mu \nu }F_{\mu \nu }$ (${\cal O}(p^{6})$ ) compared to the $%
{\cal O}(p^{4})$ $\pi \pi $-loop contribution \cite{DEG}. This prediction is
in agreement with experiment \cite{PDG98}.

$B$-physics will test SM measuring the CKM triangle \cite{stone} with sizes $%
V_{qb}^{*}V_{qd};$ the area of this triangle, $J_{CP}/2,$ is invariant for
all CKM\ triangles and non-zero if $CP$ is violated; in the Wolfenstein
parametrization:\ 
\begin{equation}
\left| J_{CP}\right| \stackrel{Wolfenstein}{\simeq }A^{2}\lambda ^{6}\eta
\label{eq:j}
\end{equation}
with $V_{us}=\lambda ,$ $V_{cb}=$ $A\lambda ^{2},\Im m(V_{td})=-A\lambda
^{3}\eta $. As a consequence of the improved understanding of low energy
physics we can test precisely (\ref{eq:j}) in rare kaon channels\cite
{reviews}.We can establish that $K\rightarrow \pi \nu \overline{\nu }$ is
dominated by far by short distance 
\begin{equation}
A(s\rightarrow d\nu \overline{\nu })_{SM}\sim
G_{F}^{2}\sum_{q=u,c,t}V_{qs}^{*}V_{qd}\quad m_{q}^{2\quad }\left( \overline{%
s}d\right) _{V-A}\left( \overline{\nu }\nu \right) _{V-A}\qquad
\label{ampsd}
\end{equation}
and due to the large top mass, is very sensitive to the CKM matrix element,$%
V_{td}$ \cite{gino98a,buchalla96}. Thus $K$-physics will measure $J_{CP}$
from the triangle with sizes $V_{qs}^{*}V_{qd}$ and test extensions of the
SM. $K$-physics is also an ideal window to test Flavour violation \cite
{kettel}: scales of 100\ TeV are probed \cite{reviews}. Here we first study $%
K\rightarrow \pi \nu \overline{\nu },$ then $K\rightarrow \pi e\overline{e},$
$K_{L}\rightarrow \mu \overline{\mu }$ and $K\rightarrow \pi \pi \gamma .$

\section{$K\rightarrow \pi \nu \overline{\nu }$}

\subsection{Standard Model}

The SM predicts the $V-A\otimes V-A$ effective hamiltonian in (\ref{ampsd})
and Wilson coefficients known at next--to--leading order \cite{buchalla96}%
.\thinspace $SU(2)$ isospin symmetry relates hadronic matrix elements for $%
K\rightarrow \pi \nu \overline{\nu }$ to $K\rightarrow \pi l\overline{\nu }$
to a very good precision \cite{Marciano96}$.$ QCD corrections have been
evaluated at next-to-leading order\cite{reviews} and the main uncertainties
is due the strong corrections to the charm loop contribution.

The structure in $($\ref{ampsd}) leads to a pure CP violating contribution
to $K_{L}\rightarrow \pi ^{0}\nu \overline{\nu },$ induced only from the top
loop contribution and thus proportional to $\Im m(\lambda _{t})$ and free of
hadronic uncertainties. This leads to the prediction \cite{buchalla96}

\begin{equation}
B(K_{L}\rightarrow \pi ^{0}\nu \overline{\nu })_{SM}=4.25\times
10^{-10}\left[ \frac{\bar{m}_{t}(m_{t})}{170GeV}\right] ^{2.3}\left[ \frac{%
\Im m(\lambda _{t})}{\lambda ^{5}}\right] ^{2}.  \label{eq:klpi0nunu}
\end{equation}

$K^{^{\pm }}\rightarrow \pi ^{\pm }\nu \overline{\nu }$ receives CP
conserving and violating contributions proportional to $\Re e(\lambda _{c}),$
$\Re e(\lambda _{t})$ and $~~\Im m(\lambda _{t}).\;$Theoretical uncertainty
from the charm loop induces 5\% error on the width. If one takes into
account the various indirect limits, i.e.$V_{ub}$ and\ $\varepsilon ,$ on
CKM elements one obtains the SM values\cite{reviews,buchalla96}: 
\begin{equation}
B(K_{L}\rightarrow \pi ^{0}\nu \overline{\nu })\ =\left( 2.8\pm 1.1\right)
\times 10^{-11}\quad B(K^{\pm }\rightarrow \pi ^{\pm }\nu \overline{\nu }%
)=\left( 0.8\pm 0.3\right) \times 10^{-10}
\end{equation}
Using only the experimental knowledge on $B_{s,d}-$ mixing the more general
bound \cite{buchalla99b} $B(K^{\pm }\rightarrow \pi ^{\pm }\nu \overline{\nu 
})_{SM}<1.67\times 10^{-10}$ is obtained. To be compared respectively with
the experimental results in \cite{kpi0pi0nunu} and in \cite{kpipnunu,kettel} 
\begin{equation}
B(K_{L}\rightarrow \pi ^{0}\nu \overline{\nu })\ \leq 5.9\times 10^{-7}{\
\qquad }B(K^{^{\pm }}\rightarrow \pi ^{\pm }\nu \overline{\nu })=\left(
1.5_{-1.2}^{+3.4}\right) \times 10^{-10}
\end{equation}
A future measurement \cite{kettel} of \thinspace $B(K_{L}\rightarrow \pi
^{0}\nu \overline{\nu })$ with 10 \% error implies a determination of $\Im
m(\lambda _{t})$, as we can see from (\ref{eq:klpi0nunu}), and the area $J$
in (\ref{eq:j}), with 5 \% error.

\subsection{New Physics}

The isospin structure of any $\overline{s}d$ operator (bilinear in the quark
fields) leads to the model independent relation \cite{GNir97} and to an
interesting bound with E787 \cite{kpipnunu,kettel} 
\[
B(K_{L}\rightarrow \pi ^{0}\nu \overline{\nu })<\frac{\tau _{K_{L}}}{\tau
_{K^{+}}}\ B(K^{^{\pm }}\rightarrow \pi ^{\pm }\nu \overline{\nu }) 
\begin{array}{c}
< \\ 
E787
\end{array}
2.9\cdot 10^{-9}\quad {\rm at\quad }90\%C.L. 
\]
Lately it has been pointed out the possibility of new physics to enhance
substantially the SM\ predictions \cite{gino98b}. Also the recent value of $%
\varepsilon ^{\prime }/\varepsilon $ \cite{KTeVeps99,Na48eps99,bertolini},
though not incompatible with the SM, allows large values for new sources of
CP violating contributions. Adopting a supersymmetric framework (susy) where
the flavour breaking scale is much larger than susy breaking scale ($M_{s})$
, which is in turn larger than the EW\ scale ($M_{s}\geq M_{W})$ the bigger
contributions to the $K\rightarrow \pi l\overline{l}$ effective hamiltonian
come from operators of dimensions four \ and five.\ Indeed once the heavy ($%
M_{s})$ degrees of freedom have been integrated out , $Z-$(susy)penguin
effects could be parametrized by an effective dimension-4 operator $%
\overline{s}dZ$ with coupling $Z_{ds}$ and with a contribution to (\ref
{ampsd}) ${\cal O(}G_{F}Z_{ds})$ \cite{gino98b,reviews}, while the superbox
diagrams generate a suppressed ${\cal O(}1/M_{s}^{2})$ effective dimension-6
operator $\overline{s}d\overline{\nu }\nu $ \cite{buras99}. We can
parametrize the relevant dimension-5 effective hamiltonian as \cite{buras99} 
\begin{equation}
{\cal H}_{eff}^{|\Delta S|=1;d=5}={\frac{4G_{F}}{\sqrt{2}}}\left[ C_{\gamma
}^{+}O_{\gamma }^{+}+C_{\gamma }^{-}O_{\gamma
}^{-}+C_{g}^{+}O_{g}^{+}+C_{g}^{-}O_{g}^{-}+\mbox{\rm h.c.}\right] ,
\label{eq:d5}
\end{equation}
\begin{equation}
O_{\gamma }^{\pm }=\frac{Q_{d}e}{16\pi ^{2}}(\bar{s}_{L}\sigma _{\mu \nu
}d_{R}\pm \bar{s}_{R}\sigma _{\mu \nu }d_{L})F^{\mu \nu }  \label{eq:d5f}
\end{equation}
\begin{equation}
O_{g}^{\pm }=\frac{g}{16\pi ^{2}}(\bar{s}_{L}\sigma _{\mu \nu
}t^{a}G_{a}^{\mu \nu }d_{R}\pm \bar{s}_{R}\sigma _{\mu \nu }t^{a}G_{a}^{\mu
\nu }d_{L})  \label{eq:d5G}
\end{equation}
In the SM\ the coefficients are chirally suppressed \cite{bertolinipenguin},
and thus extensions of SM might be interesting \cite{Masiero}.\ If we
consider the bounds coming from $\varepsilon ^{\prime }/\varepsilon $ and $%
K_{L}\rightarrow \mu \overline{\mu }$ \cite{BS99} , as we shall see, large
NP enhancements can be obtained for $K_{L}\rightarrow \pi ^{0}\nu \overline{%
\nu }$ and for $K^{^{\pm }}\rightarrow \pi ^{\pm }\nu \overline{\nu }$
(respectively at most a factor $10$ and 3 in the branching compared to the
SM\ predictions) \cite{buras99}.

\section{$K_{L}\rightarrow \pi ^{0}e^{+}e^{-}$}

Electromagnetic interactions in $K_{L}\rightarrow \pi ^{0}e^{+}e^{-}$ add
new structures to the direct CP violating contribution analogous to (\ref
{ampsd}) (additional single photon exchange contributions are smaller): i)
indirect $CP$ --violating contribution $K_{L}=K_{2}+\tilde{\varepsilon}K_{1}%
\stackrel{\tilde{\varepsilon}K_{1}}{\rightarrow }\pi ^{0}\stackrel{*}{\gamma 
}\rightarrow \pi ^{0}e^{+}e^{-}$ and ii) a $CP-$ conserving contribution: $%
K_{L}\rightarrow \pi ^{0}\stackrel{*}{\gamma }\stackrel{*}{\gamma }%
\rightarrow \pi ^{0}e^{+}e^{-}.$ Thus to really probe the short distance
window one has to have under control i) and ii) and also we must warn about
the danger of the potentially large background contribution from $%
K_{L}\rightarrow e^{+}e^{-}\gamma \gamma $ to $K_{L}\rightarrow \pi
^{0}e^{+}e^{-}$ \cite{greenlee}$.$ The present bounds from KTeV are \cite
{kettel}: 
\begin{equation}
B(K_{L}\rightarrow \pi ^{0}e^{+}e^{-})<5.6\times 10^{-10}\quad
B(K_{L}\rightarrow \pi ^{0}\mu ^{+}\mu ^{-})<3.4\times 10^{-10}
\end{equation}
and the prediction for the direct CP violation contribution analogous to (%
\ref{eq:klpi0nunu}) is 
\[
B(K_{L}\rightarrow \pi ^{0}e^{+}e^{-})_{CPV-dir}^{SM}\ =0.69\times
10^{-10}\left[ \frac{\bar{m}_{t}(m_{t})}{170GeV}\right] ^{2}\left[ \frac{\Im
m(\lambda _{t})}{\lambda ^{5}}\right] ^{2}; 
\]
using the present constrains on $\Im m(\lambda _{t})$ one obtains \cite
{reviews,BL94} 
$$2.8\times 10^{-12}\leq B(K_{L}\rightarrow \pi
^{0}e^{+}e^{-})_{CPV-dir}^{SM}\leq 6.5\times 10^{-12}.$$ 
New physics
contributions \cite{buras99} from an effective dimension-4 $Z_{ds}$ vertex
and ${\cal H}_{eff}^{|\Delta S|=1;d=5}$ in (\ref{eq:d5}) can enhance the
branching up to a factor 10.

\subsection{Indirect CP violation contribution, $K_{S}\rightarrow \pi
^{0}e^{+}e^{-}$ and $K^{\pm }\rightarrow \pi ^{\pm }l^{+}l^{-}$}

Electromagnetic interaction induce a new operator to the effective
hamiltonian for $s\rightarrow dl^{+}l^{-}:$ $\left( \overline{s}d\right)
_{V-A}\left( l^{+}l^{-}\right) _{V}$, which generates a contrbution to $%
K\rightarrow \pi l^{+}l^{-}$ dominated by long distance and thus studied in $%
\chi $PT. These decays ($K^{\pm }\rightarrow \pi ^{\pm }\stackrel{*}{\gamma }
$ and $K_{S}\rightarrow \pi ^{0}\stackrel{*}{\gamma }$ ) start at ${\cal O(}%
p^{4})$ with loops (dominated by the $\pi \pi -$cut$)$ and unknown
counterterm contributions \cite{EPR1}.\ Higher order contributions (${\cal O(%
}p^{6}))$ might be large, but not completely under control since new and
with unknown coefficients counterterm structures will appear \cite{r2}.\
However we can still parameterize, quite generally the $K\rightarrow \pi 
\stackrel{*}{\gamma }(q)$ form factor as \cite{r3} 
\begin{equation}
W_{i}(z)\,=\,G_{F}M_{K}^{2}\,(a_{i}\,+\,b_{i}z)\,+\,W_{i}^{\pi \pi
}(z)\;,\qquad \qquad i=\pm ,S  \label{eq:ctkpg}
\end{equation}
with $z=q^{2}/M_{K}^{2}$, and where $W_{+}^{\pi \pi }(z)$ is the loop
contribution, given by the $K\rightarrow \pi \pi \pi $ unitarity cut and
completely known up to ${\cal O(}p^{6})$. All our results can be expressed
in terms of the unknown parameters $a_{i}$ and $b_{i},$ expected of ${\cal O(%
}1)$. At the first non-trivial order, ${\cal O(}p^{4}),$ $b_{i}=0,$ while $%
a_{i}$ receive counterterm contributions not determined yet. At ${\cal O(}%
p^{6}),$ $b_{i}\neq 0,$ while $a_{i}$ receive new counterterm contributions.
Due to the generality of (\ref{eq:ctkpg}) we expect that $W_{i}(z)$ is a
good approximation to the complete form factor.

From the $K^{+}\rightarrow \pi ^{+}e^{+}e^{-}$ experimental width and slope,
E865 obtains \cite{kpeeE865} 
\begin{equation}
a_{+}\,=-0.587\pm 0.010\qquad \,b_{+}=-0.655\pm 0044  \label{eq:lpkpll}
\end{equation}
Also the fit with (\ref{eq:ctkpg}), i.e. with the genuine chiral \cite
{kpeeE865} contributions $W_{i}^{\pi \pi }(z),$ is better ($\chi ^{2}$ /$%
d.o.f.$ $\sim 13.3/18)$ than just a linear slope ( $\chi ^{2}$/$d.o.f.$ $%
\sim 22.9/18),$ showing the validity of the chiral expansion.Then the
universality of the form factor in (\ref{eq:ctkpg}) is further tested by
using (\ref{eq:lpkpll}) to predict the branching $B(K^{+}\rightarrow \pi
^{+}\mu ^{+}\mu ^{-}),\;$which indeed perfectely agrees with the new
experimental value\cite{kpmumuE865} $B(K^{+}\rightarrow \pi ^{+}\mu ^{+}\mu
^{-})_{{\rm {exp}}}=(9.22\pm 0.60\pm 0.49)\cdot 10^{-8}$. This value is
however larger \cite{r3} by $3.3$ $\sigma $ 's than the old experimental
result \cite{r4}. Also the slope in the muon channel, though with large
statistical errors, is now consistent with (\ref{eq:lpkpll}).

We should stress that it is not clear at the moment the meaning of the
apparent slow convergence in the chiral expansion in $K^{+}\rightarrow \pi
^{+}l^{+}l^{-},$ indeed the values in (\ref{eq:lpkpll}) do not respect the
naive chiral dimensional analysis expectation $\,b_{+}/a_{+}\sim
M_{K}^{2}/M_{V}^{2}.\qquad $

There is no model independent relation among $a_{S}$ and $a_{+}$ and thus a
secure determination of $B(K_{L}\rightarrow \pi
^{0}e^{+}e^{-})_{CP-indirect} $ requires a direct measurement of $%
B(K_{S}\rightarrow \pi ^{0}e^{+}e^{-}),$ possibly to be performed by KLOE at
DA$\Phi $NE \cite{r3}. The dependence from $b_{S}$ is very mild and thus we
predict $B(K_{S}\rightarrow \pi ^{0}e^{+}e^{-})\,\simeq
\,5.2\,a_{S}^{2}\,\times 10^{-9}~$. We can take advantage of the
interference term among direct and indirect the $CP$ --violating terms \cite
{r3}: for $a_{S}\stackrel{<}{_{\sim }}-0.5$ or $a_{S}\stackrel{>}{_{\sim }}%
1.0$ we obtain $B(K_{L}\rightarrow \pi ^{0}e^{+}e^{-})_{CPV}\stackrel{>}{%
_{\sim }}10^{-11}$ and thus perform an independent measurement of $\Im
m\lambda _{t}$, with a precision increasing with the value of $|a_{S}|$.

\subsection{CP conserving contributions: ``$\gamma \gamma "$ intermediate
state contributions}

\hspace*{0.1cm} The general amplitude for $K_{L}(p)\rightarrow \pi
^{0}\gamma (q_{1})\gamma (q_{2})$ can be decomposed as a sum of two
independent Lorentz and gauge invariant amplitudes: $A(z,y),$ where the two
photons are in a state of total angular momentum $J=0$ ($J,$ total diphoton
angular momentum), and the higher angular momentum state $B(z,y)$, also
chirally and kinematically suppressed \cite{DI98}. The double differential
rate is given by 
\begin{equation}
\frac{\displaystyle \partial ^{2}\Gamma }{\displaystyle \partial y\,\partial
z}=\,\frac{\displaystyle m_{K}}{\displaystyle 2^{9}\pi ^{3}}[%
\,z^{2}\,|\,A\,+\,B\,|^{2}\,+\,\left( y^{2}-\frac{\displaystyle \lambda
(1,r_{\pi }^{2},z)}{\displaystyle 4}\right) ^{2}\,|\,B\,|^{2}\,]~,
\label{eq:doudif}
\end{equation}
where $y=p\cdot (q_{1}-q_{2})/m_{K}^{2}$ , $z\,=%
\,(q_{1}+q_{2})^{2}/m_{K}^{2},$ $r_{\pi }=m_{\pi }/m_{K}$ and $\lambda
(a,b,c)$ is the usual kinematical function. Thus in the region of small $z$
(collinear photons) the $B$ amplitude is dominant and can be determined
separately from the $A$ amplitude. This feature is crucial in order to
disentangle the CP-conserving contribution $K_{L}\rightarrow \pi
^{0}e^{+}e^{-}.$ Here we assume, consistentely with models \cite{r2} and
possibly to test in $K_{L}\rightarrow \pi ^{0}e^{+}e^{-}\gamma $, that the
dominant contribution is generated by on-shell $\gamma \gamma $ rescattering 
\cite{Gabbiani97}. It turns out that for the $J=0$ contribution $%
A(K_{L}\rightarrow \pi ^{0}e^{+}e^{-})_{J=0}\sim m_{e}$ ($m_{e}$ electron
mass) \cite{EPR88}; while the suppressed $B$-type amplitude, generate $%
A(K_{L}\rightarrow \pi ^{0}e^{+}e^{-})_{J\neq 0}$ competitive with the CP
violating contributions \cite{r2}.

The leading finite ${\cal O}(p^{4})$ amplitudes of $K_{L}\rightarrow \pi %
^{0}\gamma \gamma $ generates only the $A$--type amplitude in Eq.~(\ref
{eq:doudif}) \cite{cappiello}, but underestimates the observed branching
ratio \cite{KTeVkpgg99,kettel}, $(1.68\pm 0.07\pm 0.08)\times 10^{-6}$ by a
large factor. The two presumably large ${\cal O}(p^{6})$ contributions have
been studied: i) ${\cal O}(p^{6})$ unitarity corrections \cite{CD93,CE93}
enhance the ${\cal O}(p^{4})$ branching ratio by $40\%$, and generate a $B$%
--type amplitude, ii) vector meson exchange contributions to $%
K_{L}\rightarrow \pi ^{0}\gamma \gamma $ are in general model dependent \cite
{SE88,EP90} but one can parameterize their contribution to $A$ and $B$ by an
effective vector coupling $a_{V}$ \cite{EP90} . The agreement with
experimental $K_{L}\rightarrow \pi ^{0}\gamma \gamma $ rate and spectrum
would demand $a_{V}\sim -0.8\;$\cite{CE93,DP97} . We have related $a_{V}$
with the linear slope, $\alpha ,$ of the $K_{L}\rightarrow \gamma \gamma ^{*}
$ form factor \cite{DP97}, which is also generated by vector meson exchange
contribution$.\;$In factorization it is possible to describe VMD
contributions to these observables in terms of one free\ parameter $k_{F}$.\
The resulting phenomenology is successful and it is also suggested that very
interestingly matching with short distance should be performed at the
resonance scale and not at the kaon mass \cite{DP97,DP98}. The new data from
KTeV \cite{KTeVkpgg99} confirms sharply our prediction $:$ $a_{V}=-0.72\pm %
0.05\pm 0.06$ and show a clear evidence of events at low z. This turns in a
stringent determination for the CP conserving contribution to $%
K_{L}\rightarrow \pi ^{0}e^{+}e^{-}$: $1.<B(K_{L}\rightarrow \pi %
^{0}e^{+}e^{-})\cdot 10^{12}<4$ \cite{r2,DP97}.

\section{$K_{L}\rightarrow l^{+}l^{-}$}

To fully exploit the potential of $K_{L}\rightarrow \mu ^{+}\mu ^{-}$ in
probing short--distance dynamics it is necessary to have a reliable control
on its long--distance amplitude. However the dispersive contribution
generated by the two--photon intermediate state cannot be calculated in a
model independent way and it is subject to various uncertainties \cite
{DIP,V98,gomez}. The branching ratio can be generally decomposed as $%
B(K_{L}\rightarrow \mu ^{+}\mu ^{-})\,=\,|{\Re e{\cal A}}|^{2}\,+\,|{\Im m%
{\cal A}}|^{2}$, and the dispersive contribution can be rewritten as ${\Re e%
{\cal A}}\,=\,{\Re e{\cal A}}_{long}\,+\,{\Re e{\cal A}}_{short}$. The
recent measurement of $B(K_{L}\rightarrow \mu ^{+}\mu ^{-})=\,(7.18\pm
0.17)\times 10^{-9}$ \cite{PDG98} is almost saturated by the absorptive
contribution $B(K_{L}\rightarrow \mu ^{+}\mu ^{-})_{abs}=(7.1\pm 0.2)\times
10^{-9}$, obtained from the experimental $K_{L}\rightarrow \gamma \gamma $
width. This leaves very small room for the dispersive contribution
determined by $B(K_{L}\rightarrow \mu ^{+}\mu ^{-})_{{\rm exp}}$,~${\Re e%
{\cal A}}_{exp}=\,(-1.0\pm 3.7)\times 10^{-10}$ or $|{\Re e{\cal A}}%
_{exp}|^{2}\,<\,5.6\times 10^{-10}$ at $90\%$ C.L.\cite{DIP}.

Within the Standard Model the known\cite{BB} NLO short-distance contribution 
$|{\Re e{\cal A}}_{short}|^{2}=\,(0.9\pm 0.4)\times 10^{-9}$, gives the
possibility to extract information on $V_{td}$ and New Physics, once ${\Re e%
{\cal A}}_{long}$ is under control. Two methods \cite{DIP,gomez} and a
criticism \cite{V98} have been proposed to predict ${\Re e{\cal A}}_{long}$%
.\ 

Large-$N_{c}$ argument and $U(3)\otimes U(3)$ symmetry have been invoked in 
\cite{gomez} for ${\Re e{\cal A}}_{long};$ however these arguments give a
poor phenomenological description of the experimental $K_{L}\rightarrow
\gamma \gamma $ width and thus caution must be used before to completely
accept this result \cite{derafael}.

We instead have proposed \cite{DIP} a low energy parameterization of the $%
K_{L}\rightarrow \gamma ^{*}\gamma ^{*}$ form factor that include the poles
of the lowest vector meson resonances with arbitrary residues 
\begin{equation}
f(q_{1}^{2},q_{2}^{2})=1+\,\alpha \left( \frac{\displaystyle q_{1}^{2}}{%
\displaystyle q_{1}^{2}-m_{V}^{2}}+\frac{\displaystyle q_{2}^{2}}{%
\displaystyle q_{2}^{2}-m_{V}^{2}}\right) \,+\,\beta \,\frac{\displaystyle %
q_{1}^{2}q_{2}^{2}}{\displaystyle (q_{1}^{2}-m_{V}^{2})(q_{2}^{2}-m_{V}^{2})}%
~.  \label{eq:fq1q2}
\end{equation}
The parameters $\alpha $ and $\beta $, expected to be ${\cal O}(1)$ by
chiral power counting, are in principle directly accessible by experiment in 
$K_{L}\rightarrow \gamma \ell ^{+}\ell ^{-}$ and $K_{L}\rightarrow
e^{+}e^{-}\mu ^{+}\mu ^{-}$. Now we can determine only  $\alpha _{exp},$
while $\beta $ is$\ $constrained by QCD: the form factor defined in Eq.~(\ref
{eq:fq1q2}) goes as $1+2\alpha +\beta $ for $q_{i}^{2}\gg m_{V}^{2}$ and one
has to introduce an ultraviolet cutoff $\Lambda $; since in this region the
perturbative QCD calculation of the two-photon contribution gives a small
result we find $|1+2\alpha +\beta |\ln (\Lambda /M_{V})<0.4$, showing  a
mild behaviour of the form factor at large $q^{2}$ and limiting $\beta $ .
Thus we predict \cite{DIP} $|{\Re e{\cal A}}_{long}|<2.9\times %
10^{-5}\;\;\;\;(90\%C.L.)$. These bounds could be very much improved if the $%
\alpha $ and $\beta $ parameters were measured with good precision and a
more stringent bound on $|{\Re e{\cal A}}_{exp}|$ is established. Recentely $%
K_{L}\rightarrow e^{+}e^{-}$ has been measured at BNL E871 \cite{ambrose}:\ $%
B(K_{L}\rightarrow e^{+}e^{-})=(8.7_{-4.1}^{+5.7})\times 10^{-12},$ however
the theoretical prediction\cite{V98,gomez} for this branching is not
sensitive to the slopes of the form factor.

\section{$K\rightarrow \pi \pi \gamma $}

The $K\rightarrow \pi \pi \gamma $ amplitude is usually decomposed also in
electric ($E)$ and the magnetic ($M)$ terms \cite{ecker94,DI98,DMS93}. In
the electric transitions one generally separates the bremsstrahlung
amplitude $E_{B}$ generally enhanced from the factor ${1/}E_{\gamma }^{*}$
for $E_{\gamma }^{*}\rightarrow 0,$ ($E_{\gamma }^{*}$ is the photon energy
in the kaon rest frame). Summing over photon helicities there is no
interference among electric and magnetic terms. At the lowest order ($p^{2})$
in $\chi PT$ one obtains only $E_{B}.$ Magnetic and electric direct emission
amplitudes, appearing at ${\cal O(}p^{4}),$ can be decomposed in a multipole
expansion \cite{LV88,DMS93,DI98}. We do not discuss $K_{S,L}\to \pi ^{0}\pi %
^{0}\gamma $ due to the small branching ratio ($<10^{-8})$ \cite{HS93} and $%
K_{S}\to \pi ^{+}\pi ^{-}\gamma $ \cite{DMS}, where no new experimental
results have been reported recentely. While motivated by new results we
update $K_{L}\rightarrow \pi ^{+}\pi ^{-}\gamma $ and $K^{+}\rightarrow \pi %
^{+}\pi ^{0}\gamma .$

 $
\begin{array}[t]{|c|c|c|}
\hline
\mbox{decay} & BR(\mbox{bremsstrahlung}) & BR(\mbox{direct 
emission}) \\ \hline
\begin{array}{l}
K^{\pm }\to \pi ^{\pm }\pi ^{0}\gamma \\ 
T_{\pi ^{+}}^{*}=(55-90)MeV
\end{array}
& 
\begin{array}{c}
(2.57\pm 0.16)\times 10^{-4} \\ 
(\Delta I=3/2)
\end{array}
& 
\begin{array}{c}
(4.72\pm 0.77)\times 10^{-6} \\ 
E1,M1
\end{array}
\\ \hline
\begin{array}{l}
K_{L}\to \pi ^{+}\pi ^{-}\gamma \\ 
E_{\gamma }^{*}>20MeV
\end{array}
\quad & 
\begin{array}{c}
(1.49\pm 0.08)\times 10^{-5} \\ 
(CP{\rm {\ violation)}}
\end{array}
& 
\begin{array}{c}
(3.09\pm 0.06)\times 10^{-5} \\ 
M1,E2
\end{array}
\\ \hline
\end{array}$

\vskip 0.5cm

\underline{$K_{L}\rightarrow \pi ^{+}\pi ^{-}\gamma $}($\gamma ^{*})$:
bremsstrahlung ($E_{B})$ is suppressed by $CP$ violation\ ($\eta _{+-}$) and
firmly predicted theoretically \cite{DMS93}. This contribution has been also
measured by interference with the $M1$ transition in $K_{L}\rightarrow \pi %
^{+}\pi ^{-}e^{+}e^{-}$ \ \cite{belz99,Yamanaka}. Due to the large slope,
KTeV parametrizes the magnetic amplitude 

\begin{equation}
{\cal F=}\widetilde{g}_{M1}\left[ \frac{a_{1}}{(M_{\rho
}^{2}-M_{K}^{2})+2M_{K}E_{\gamma }^{*}}+a_{2}\right]
\label{amplitudeKLpipig}
\end{equation}
finding $a_{1}/a_{2}=(-0.729\pm 0.026($stat))$GeV^{2}$ and the branching
given in the table, which fixes also $\widetilde{g}_{M1}$.\ Such large slope
can be accomodated in various Vector dominance schemes \cite
{HS93,ecker94,kppgVMD}, while the rate is very sensitive to $SU(3)-$breaking
and unknown $p^{4}$ unknown low energy contributions and thus difficult to
predict.

\underline{$K^{+}\rightarrow \pi ^{+}\pi ^{0}\gamma $} New data from BNL\
E787\cite{Komatsubara99} show vanishing interference among
brems\-strah\-lung and elec\-tric tran\-sition. Thus the di\-rect emis\-sion
bran\-ching ($B(K^{+}\rightarrow \pi ^{+}\pi ^{0}\gamma $)$_{{\rm {exp}}%
}^{DE}),$ in the table, must be interpreted as a pure magnetic transition.
Theoretically one can identify two different sources for $M,$ appearing at $%
{\cal O(}p^{4}):$ i) a pole diagram with a Wess-Zumino term and ii) a pure
weak contact term, also generated in factorization by an anomalous current 
\cite{CH90a}. \ $B(K^{+}\rightarrow \pi ^{+}\pi ^{0}\gamma $)$_{{\rm {\ exp}}%
}^{DE}$ is substantially smaller than previous values, but still show that
this last contribution is non-vanishing.

{\bf CP\ Violation}

Direct $CP$ violation can be established in the width charge asymmetry in $%
K^{\pm }\rightarrow \pi ^{\pm }\pi ^{0}\gamma ,$ $\delta \Gamma /2\Gamma $
and \ in the interference $E_{B}$ with $E_{1}$ in $K_{L}\rightarrow \pi ^{+}%
\pi ^{-}\gamma $ ($E$ with $M_{1}$ in $K_{L}\rightarrow \pi ^{+}\pi %
^{-}e^{+}e^{-})$; both observables are kinematically difficult since one is
looking for large photon energy distribution\cite{DI98}. SM charge asymmetry
were looked in \cite{CPold} expecting $\delta \Gamma /2\Gamma \leq 10^{-5}$.
General bounds on new physics in $M1$ transitions have been studied in \cite
{HV99}, while the effects of dimension-5 operators in (\ref{eq:d5}) $E1$
transitions have been studied in \cite{CIP}, where for instance it has been
shown that the value of $\Re \left( \varepsilon ^{\prime }/\varepsilon
\right) $ allows in particular kinematical regions a factor 10\ larger than
SM.

\section{Conclusions}

The recent experimental results in $K-$decays, i.e. $\varepsilon ^{\prime
}/\varepsilon ,$ $K\rightarrow \pi \nu \overline{\nu }$ and $%
K_{L}\rightarrow \pi ^{0}\gamma \gamma ,$ let us hope that the near future
will be full of exciting new results and challenges to the SM and its
extensions. In particular $K\rightarrow \pi \nu \overline{\nu }$ can
determine $J$ in (\ref{eq:j}). I remind that, though not discussed here,
T-invariance \cite{Yamanaka} and Flavour violation\cite{kettel} are other
interestiong windows to search for new physics. From the theoretical side we
expect an improving in matching long and short distance contributions\cite
{derafael} and this would lead to an accurate determination of low energy
parameters, for instance a reliable relation between $a_{S}$ and $a_{+}$ in (%
\ref{eq:ctkpg}).

\section{Acknowledgements}

I am happy to thank the organizers of the Conference for the nice atmosphere
and G.Isidori for many interesting discussions. This work is supported in
part by TMR, EC--Contract No. ERBFMRX--CT980169 (EURODA$\Phi $NE).

\end{document}